\documentclass[preprint,12pt,authoryear]{elsarticle}

\usepackage{amssymb}

\journal{Planetary end Space Science}

\begin{document}

\begin{frontmatter}



\title{Meteoroid Orbits from Video Meteors\\The Case of the Geminid Stream}
\author[label1]{M\'{a}ria Hajdukov\'{a} Jr.}\ead{Maria.Hajdukova@savba.sk}
\author[label2]{Pavel Koten}
\author[label3]{Leonard Korno\v{s}}
\author[label3]{Juraj T\'{o}th}
\address[label1]{Astronomical Institute of the Slovak Academy of Sciences, Bratislava, Slovakia}
\address[label2]{Astronomical Institute of the Czech Academy of Sciences, Ond\v{r}ejov, Czech Republic}
\address[label3]{Faculty of Mathematics, Physics and Informatics, Comenius University, Bratislava,
Slovakia}


\begin{abstract}
We use the Slovak and Czech video meteor observations, as well as
video meteoroid orbits collected in the CAMS, SonotaCo, EDMOND and
DMS catalogues, for an analysis of the distribution of meteoroid
orbits within the stream of the Geminids and of the dispersion of
their radiants. We concentrate on the influence of the measurement
errors on the precision of the orbits obtained from the video
networks that are based on various meteor-detection software
packages and various meteor orbital element softwares.

The Geminids radiant dispersion obtained from the large video
catalogues reaches the dispersion of the radio observed Geminids,
wherby the diffused marginal regions are affected mostly by
meteoroids with extreme values (small or large) of the semi-major
axes. Meteoroids of shorter semi-major axes concentrate at the
eastern side of the radiant area and those of longer semi-major
axes at the western part.

The observed orbital dispersions in the Geminid stream described
by the median absolute deviation range from 0.029 to 0.042
$\mathrm{AU^{-1}}$ for the video catalogues. The distribution of
the semi-major axes of video meteors in all the databases, except
for the Ond\v{r}ejov (Czech) data, seem to be systematically
biased in comparison with the photographic and radio meteors. The
determined velocities of the video data are underestimated,
probably as a consequence of the methods used for the positional
and velocity measurements. The largest shift is observed in the
EDMOND and SonotaCo catalogues.

Except for the measurement errors which influence the analyses and
their interpretations, we also point out the problem of the
uncertainties of the numerical integration procedures that
influence the simulations' results. Several experimental
integrations of the Geminids parent asteroid, which we performed
from the present to the past and then back to the year 2015,
showed that a complete reproduction, including also the mean
anomaly, is only possible for a time span of about 2700 years.

\end{abstract}


\end{frontmatter}

\section{Introduction}

\label{} The recent rapid development of video techniques is
reflected in the massive increase in detected meteors. This is of
high significance for minor meteor showers radiants
determinations, derivations of meteoroid flux densities and other
purposes. However, the production of a large number of meteor
orbits often comes at the expense of their quality. This is then
reflected in the meteor's characteristics and influences further
analyses, models of the meteoroid streams and searches for the
parent bodies. The biggest problem is the measurement and
determination of velocity, as the value of the semi-major axis is
very sensitive to the value of the heliocentric velocity. When
studying the structure of meteoroid streams through shower
meteors, the fact that the original orbital dispersion can be
smeared by larger observational and measurement errors has also to
be considered. The initial dispersion of meteoroids in a stream is
influenced by a number of processes, which appear during different
stages of the stream evolution. An overview of the underlying
principles of meteor stream formation and evolution has been given
by \cite{kresak1992}, \cite{williams2003}, and others. The effect
of these processes on the structure of meteoroid streams naturally
depends on the type of stream \citep{williamsryabova2011}.
However, \cite{kresak1992}, analyzing widely dispersed annual
meteor showers from the photographic catalogues of the IAU MDC,
showed that the measurement errors can be two or three orders of
magnitude larger than the dispersion produced by planetary
perturbations integrated over several revolutions. The most
significant source of uncertainty in semi-major axis determination
is inaccuracy in the heliocentric velocity $\mathrm{v_H}$. Errors
in $\mathrm{v_H}$ of $\mathrm{1\,km/s}$ correspond to about
$\mathrm{0.08\,to\,0.09\, AU^{-1}\,in\,1/a}$. For short-period
meteoroid streams, differences in velocity are less representative
and the dispersion in the semi-major axes is smaller. Thus,
discovering errors is more difficult because they do not produce a
spurious hyperbolicity as clear evidence of their presence, as is
the case with long-period streams \citep{hajdukova2011,
hajdukova2013}.

\label{} The Geminids is one of the largest showers in the meteor
databases, which was observed using different techniques and
studied by a large number of researchers over a large time scale.
A comprehensive review of observational and theoretical studies of
the Geminid stream was published by \cite{neslusan2015}. The
Geminid meteor shower, observed by video technique, was reported
by \cite{uedafujiwara1993, delignieetal1993, elliotetal1993,
andreicsegon2008, jenniskensetal2010, jenniskensetal2011,
jenniskensetal2016a, trigoetal2010, tothetal2011, tothetal2012,
rudawskaetal2013, madiedoetal2013, molauetal2015, molauetal2016},
and others.

In this study, we concentrate on the influence of both the
accuracy of various measurements and the precision of orbit
determination on the distribution of meteor orbits within the
stream of Geminids and on the dispersion of their radiant points.
The dispersions are studied, comparing several catalogues
(introduced in section 2), which enables the specific features of
the Geminids, as well as the diversities of the catalogues, to be
shown. In section 3, we describe the dispersion of the Geminids'
radiant points, and in section 4, the dispersion of their orbits.
We also discuss the dynamics of the Geminid stream in terms of the
uncertainties of the numerical integration procedures, which
reflect the reliability of the results obtained (section 5).

\section{Video orbits and their precision}

The necessity of high quality orbits of video meteors and
precision in their velocity measurements has been discussed by
\cite{atreyaetal2012} and \cite{egaletal2014}, who, to make
improvements, introduced the CABERNET ({\it Camera for Better
Resolution Network}) system. The importance of an error analysis
was reported by \cite{drolshagenetal2014} and \cite{albinetal2015}
in their analysis of the meteor velocity distribution from the
CILBO ({\it Canary Island Long-Baseline Observatory}) double
station video camera data. The accuracy of video meteor orbits was
discussed by \cite{skocicetal2016}, who analysed several major
showers obtained by several video networks including {\it the
Croatian Meteor Network}.

In this study, we analyze Geminids from six different video
catalogues. The individual samples of the Geminids were obtained
using the Welch procedure (\cite{welch2001}, described in more
detail in section 5). All of them fulfill the Southworth-Hawkins
D-criterion for orbital similarity \citep{southworth1963}, with
the condition $\mathrm{D_{SH}\,<\,0.2}$. The data used are
summarized in Table 1. The observed orbital and radiant
dispersions of video Geminids, including the measurement errors,
obtained separately for each catalogue, are compared with those
obtained from the photographic and radio Geminids selected from
the IAU Meteor Data Center \citep{lindbladetal2005,
neslusanetal2014, lindblad2003}.

We used data from our own video observations, carried out in the
Slovak and Czech Republics, that are collected in the Slovak Video
Meteor Network's database \citep{tothetal2015} and in the Czech
Catalogue of Video Meteor Orbits \citep{kotenetal2003}. We also
selected orbits of the Geminids from several video catalogues that
were available: the Cameras for Allsky Meteor Surveillance (CAMS)
Meteoroid Orbit Database \citep{jenniskensetal2011}, Dutch Meteor
Society Video Database \citep{delignie1996}, the SonotaCo Shower
Catalogue \citep{sonotaco2009}, and the European Video Meteor
Network Database - EDMOND \citep{kornosetal2014}. The data used
are based on various meteor-detection software packages and
various meteor orbital element softwares. In this section,
therefore, we introduce briefly all the databases of the video
networks used, their instrumentation and data reduction.

\subsection{Slovak Video Meteor Network's (SVMN) Database} \label{} The
Slovak Video Meteor Network, governed by Comenius University in
Bratislava, consists of four video stations situated in various
locations in Slovakia, which monitor meteor activity above Central
Europe. The SVMN uses the semi-automatic all sky video cameras
(All-sky Meteor Orbit System, AMOS), which record meteors of $+4$
magnitude and brighter \citep{tothetal2015}. For meteor detection
and astrometric data reduction, UFOCapture software and
UFOAnalyzer (SonotaCo, 2009) are used. For meteor orbit
computation, the new Meteor Trajectory software, based on the
\cite{ceplecha1987} paper, was developed \citep{kornosetal2015}.
The program computes orbital and geophysical parameters, together
with their uncertainties, based on the Monte Carlo simulation. The
velocity determination, giving uncertainties about
$\mathrm{0.1\,km/s}$, is still being worked on. So far, the
achieved precision is $\mathrm{<\,3\,deg}$ in radiant position and
$\mathrm{<\,10\,\%}$ in velocity. Data from SVMN are continuously
published \citep{tothetal2015} and contribute to the EDMOND
database (see section 2.5). However, the Slovak data from the SVMN
observations are also analysed differently. For the EDMOND,
velocity and radiant are calculated as average values from the
stations (according to the Sonotaco), while in the SVMN,
trajectory and velocity are calculated by our own software
\citep{kornosetal2015} based on \cite{ceplecha1987} and by fitting
the observed meteor velocity.

\subsection{Czech Catalogue of Video Meteor Orbits (Ond\v{r}ejov
data)}

The Czech database of the video meteors \citep{kotenetal2003}
contains data obtained within the double station observational
campaigns carried out in the Czech Republic. These campaigns were
dedicated to several selected meteor showers. The video cameras,
connected with image intensifiers, were aimed at one particular
meteor shower during each campaign, so the geometry of observation
was optimized for this shower. The limiting magnitude is +6. The
observed data are recorded in time resolution $0.04$ second.
Records are searched using automatic detection software MetRec
\citep{molau1999}. Found meteor images are digitalized with a PC
frame grabber, transformed into 768 x 576 pixel, 8-bit monochrome
images, and stored as sequences in a non-compressed AVI format.
All the recorded meteors are carefully reviewed and only records
of good quality are taken into account. Raw data are measured
manually and atmospheric trajectories and heliocentric orbits
calculated. No automatic reduction or calculation software is
applied. Only well proven methods are used for image measurement
\citep{koten2002}, and trajectory and orbit calculations
\citep{borovicka1990}. The errors of the measurement are
propagated through the calculation to the errors of the
parameters. The achieved precision is usually a few tenths of a
degree in the radiant position and up to $\mathrm{0.5\,km/s}$ in
the velocity. The database contains only reliable data, at the
expense of the total number of trajectories and orbits, which is
rather small.

\subsection{Dutch Meteor Society Video (DMS) Database}

The double-station video observations in the Netherlands are among
the first video observations of meteors, which started about 30
years ago. Their cameras recorded meteors with the limiting
magnitude $+7$, in a 25 degrees field of view \citep{delignie1996,
delignie1999}. The data reduction was done with the AstroRecord
measuring program and with the \cite{ceplecha1987} software for
calculating trajectories and orbital elements
\citep{deligniebetlem1999}. The double-station meteors are
measured with an astrometric accuracy of 45 arc seconds. The
results from video observation campaigns by the Dutch Meteor
Society have been published on a regular basis and are available
at their website: http://dmsweb.home.xs4all.nl/video/video.html.

\subsection{Cameras for Allsky Meteor Surveillance (CAMS) Database}

The Cameras for Allsky Meteor Surveillance system operates 60
identical narrow-angle field-of-view cameras at three locations in
California \citep{jenniskensetal2011}, detecting mostly $+4$ to $
-2$ magnitude meteors. The data are automatically processed using
the detection algorithms and modules from the MeteorScan software
package \citep{gural1995, gural1997} and a newly developed
software for calibration and multistation coincidence processing
which produces atmospheric trajectories and orbital elements
\citep{jenniskensetal2016a, jenniskensetal2016b}. The achieved
precision is $\mathrm{<\,2}$ deg in radiant direction and
$<\,10\,\%$ in velocity (mean values are $\mathrm{0.24\,deg}$ and
$\mathrm{2\,\%}$, respectively). The project was designed to
validate the unconfirmed showers in the IAU working list of meteor
showers. The meteors assigned to the various showers are
identified in the CAMS Meteoroid Orbit Database 2.0 (which can be
accessed at http://cams.seti.org.) and are submitted to the IAU
Meteor Data Center.

\subsection{European Video Meteor Network Database (EDMOND)}

The multi-national network EDMOND \citep{kornosetal2014} was
created thanks to the broad international cooperation of video
meteor observers from several European countries. The national
networks involved are: BOAM (Base des Observateurs Amateurs de
Metéores, France); BosNet (Bosnia); CEMeNt (Central European
Meteor Network, cross-border network of Czech and Slovak amateur
observers); CMN (Croatian Meteor Network or Hrvatska Meteorska
Mreza, Croatia); FMA (Fachgruppe Meteorastronomie, Switzerland);
HMN (Hungarian Meteor Network or Magyar Hullócsillagok Egyesulet,
Hungary); MeteorsUA (Ukraine); IMTN (Italian amateur observers in
Italian Meteor and TLE Network, Italy); NEMETODE (Network for
Meteor Triangulation and Orbit Determination, United Kingdom); PFN
(Polish Fireball Network or Pracownia Komet i Meteorów, PkiM,
Poland); Stjerneskud (Danish all-sky fireball cameras network,
Denmark); SVMN (Slovak Video Meteor Network, Slovakia); UKMON (UK
Meteor Observation Network, United Kingdom), BRAMON (BRAzilian
MeteOr Network), and the International Meteor Organization Video
Meteor Network (IMO VMN). Meteors (registered with the limiting
magnitude +4) are obtained and reduced using two different tools,
the MetRec \citep{molau1999} and UFO softwares
\citep{sonotaco2009}. The computation of meteor orbits is
performed by using the UFOOrbit software; therefore, data obtained
by the MetRec software are converted into the UFO format using the
SonotaCo program INF2MCSV. Multiple filters and selective criteria
involving the quality parameters (as defined in the SonotaCo
format) are applied \citep{kornosetal2013} to eliminate meteors
with the largest errors in their velocity determination. For our
analysis, we applied additional criteria to separate orbits of the
highest quality from the EDMOND data. These included: the meteor
trail had to be longer than 1 degree, the duration of the trail
had to be over 0.3 s, and the entire meteor trail had to be inside
the field of view of at least two video meteor stations.

\subsection{SonotaCo Meteor Shower Catalogue}

Video observations have been carried on for more than a decade by
the SonotaCo consortium \citep{sonotaco2009}, using more than 100
wide-angle video cameras at 25 stations in Japan. The network
registers meteors mostly up to +2 magnitude. Data are reduced
using the UFO software package developed by \cite{sonotaco2009},
which makes the catalogue homogenous in spite of the large number
of individual observers. On account of their larger fields of
view, the precision of the radiant positions measured is
approximately a factor of two less precise than the CAMS network
\citep{rudawskajenniskens2014}, which corresponds to an average
spread in radiant of about 5 deg. A detailed analysis of the
SonotaCo meteor orbits concerning the qualitative aspects was made
in the paper by \cite{verestoth2010}. The SonotaCo network
simultaneously-observed meteor data sets are freely accessible at
http://sonotaco.jp/doc/SNM/. For our analysis, in the case of the
SonotaCo data, we applied additional selective criteria, as we did
with the EDMOND database \citep{hajdukovaetal2014}, and used the
obtained subset of higher quality orbits.
\\
\\
Except for the above mentioned video networks, there are some
others we would like to mention to complete this overview. Another
automated system to observe video meteors is {\it the Canadian
Automated Meteor Observatory (CAMO)}. Their wide-field camera
system, appropriate for measurements of meteoroid fluxes, has
average radiant errors of $\mathrm{0.3\,deg}$ and speed
uncertainties of
$3\,\%$ \citep{weryketal2013, muscietal2012}.\\

{\it The Spanish Meteor Network (SPMN)} \citep{trigoetal2007,
trigoetal2008, madiedoetal2008}, which uses high-sensitivity CCD
video devices and CCD all-sky cameras, provides high accuracy
orbits of meteors, especially fireballs.

\begin{table}[t] \small
\begin{center}
\caption{\it The Geminids analysed in this work. All the orbits
from the video catalogues used were selected under the condition
$\mathrm{D_{SH}\,<\,0.2}$ of the Southworth-Hawkins D-criterion
for orbital similarity.} \label{t1} \hspace {0.5cm}
\begin{tabular}{lr}
 \hline
Database & No of Geminids \\
\hline
The Slovak Video Meteor Network (SVMN)          & 143 \\
The Czech Catalogue of Video Meteor Orbits (OND\v{R}EJOV)       & 74 \\
The Cameras for Allsky Meteor Surveillance (CAMS)       & 4827 \\
The SonotaCo Shower Catalogue (SONOTACO)            & 8264 \\
The European Video Meteor Network Database (EDMOND)         & 2401 \\
The Dutch Meteor Society Video Database (DMS)            & 104 \\
\hline
\end{tabular}
\end{center}
\end{table}

\section{Geminid shower. Radiant and speed}

\begin{figure}
\centerline{\includegraphics[width=10cm]{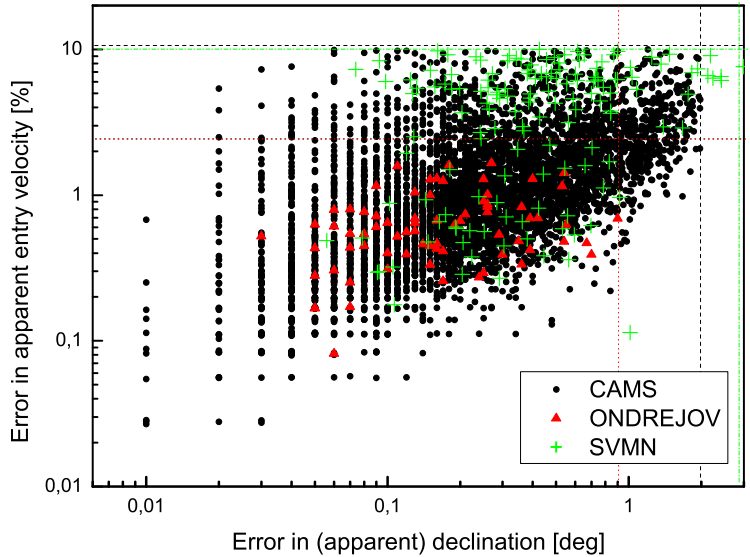}}
\caption[f2]{Error in apparent entry velocity and in declination
of the Geminids from Czech and Slovak video meteor data compared
with those from CAMS. The vertical and horizontal lines represent
the limits for errors in speed and declination for Czech (red
dotted lines), Slovak (green dash-dotted lines) and CAMS (black
dashed lines) data.} \label{FIG6}
\end{figure}

Measurement errors are, in general, smaller for long-lasting
meteors. Their distributions for Geminid meteors obtained by
Slovak and Czech video observations, and compared with CAMS data,
are shown in figure 1. The plot is based on figure 10 from
\cite{jenniskensetal2011}, who compared their mean measurement
errors to those reported for several other surveys, using various
observational techniques. The apparent radiants in the CAMS data
are measured with a median uncertainty of $\pm0.31$ deg having a
standard deviation of $\mathrm{0.42\,deg}$. Their uncertainty in
apparent entry velocity is $\mathrm{\pm0.53\,km/s}$, with a
standard deviation of $\mathrm{0.91\,km/s}$
\citep{jenniskensetal2011}. For CAMS Geminids, the median error in
the entry velocity has a value of $\mathrm{\pm0.35\,km/s}$, and in
the declination $\mathrm{\pm0.21\,deg}$. The median uncertainties
of Ond\v{r}ejov Geminids are $\mathrm{\pm0.21\,km/s}$ in entry
speed and $\mathrm{\pm0.155}\deg$  in declination. As it is seen
from figure 1, no Geminids with an error larger than
$\mathrm{0.09\,deg}$ in declination and $\mathrm{2\,\% }$ in speed
are in the Czech video catalogue, in comparison with the values of
$\mathrm{2\,deg}$ and $\mathrm{10\,\%}$ for CAMS data.

The error in declination for the Geminids observed by the Slovak
Video Meteor Network reaches up to $\mathrm{2.9\,deg}$, and the
velocity error is kept under $\mathrm{10\,\%}$. AMOS cameras have
a large field of view and high sensitivity at the same time, which
leads to the detection of faint and short meteors on one side, and
bright and long meteors, up to fireballs, on the other side. As a
result, both meteors of lower precision (due to the limitation of
a small number of observed points on the atmospheric trajectory)
and meteors of higher precision (with sufficient data points for
the velocity fit and precise trajectory determination) are
obtained. This is why two groups partly overlapping each other on
the uncertainty velocity distribution graph (figure 1) appear; the
first group with a good velocity fit with small uncertainty (under
2\% in the velocity error) and the second group with velocity
determined solely by the arithmetic mean from the first third of
the atmospheric trajectory (with the velocity error above 2\%).

\subsection{The mean radiant and its motion}

The influence of measurement errors can also be seen from the
distributions of the radiant points. \cite{kresakporubcan1970},
analysing several photographically observed major meteor showers
showed that the radiant dispersion undoubtedly increases with the
velocity dispersion, and vice versa. For most of the showers
analysed, the positional errors were negligible in comparison with
the actual deviations from the mean value. However, the authors
concluded that errors in the radiant positions may play some role
in the relatively small radiant area of the Geminid shower. The
deviations of shower meteors with respect to their position and
velocity were analysed also by \cite{molau2008}, who demonstrated
that, for short-lasting meteors, even small measurement errors can
result in large errors of their direction.

\begin{figure}
\centerline{\includegraphics[width=6.8cm]{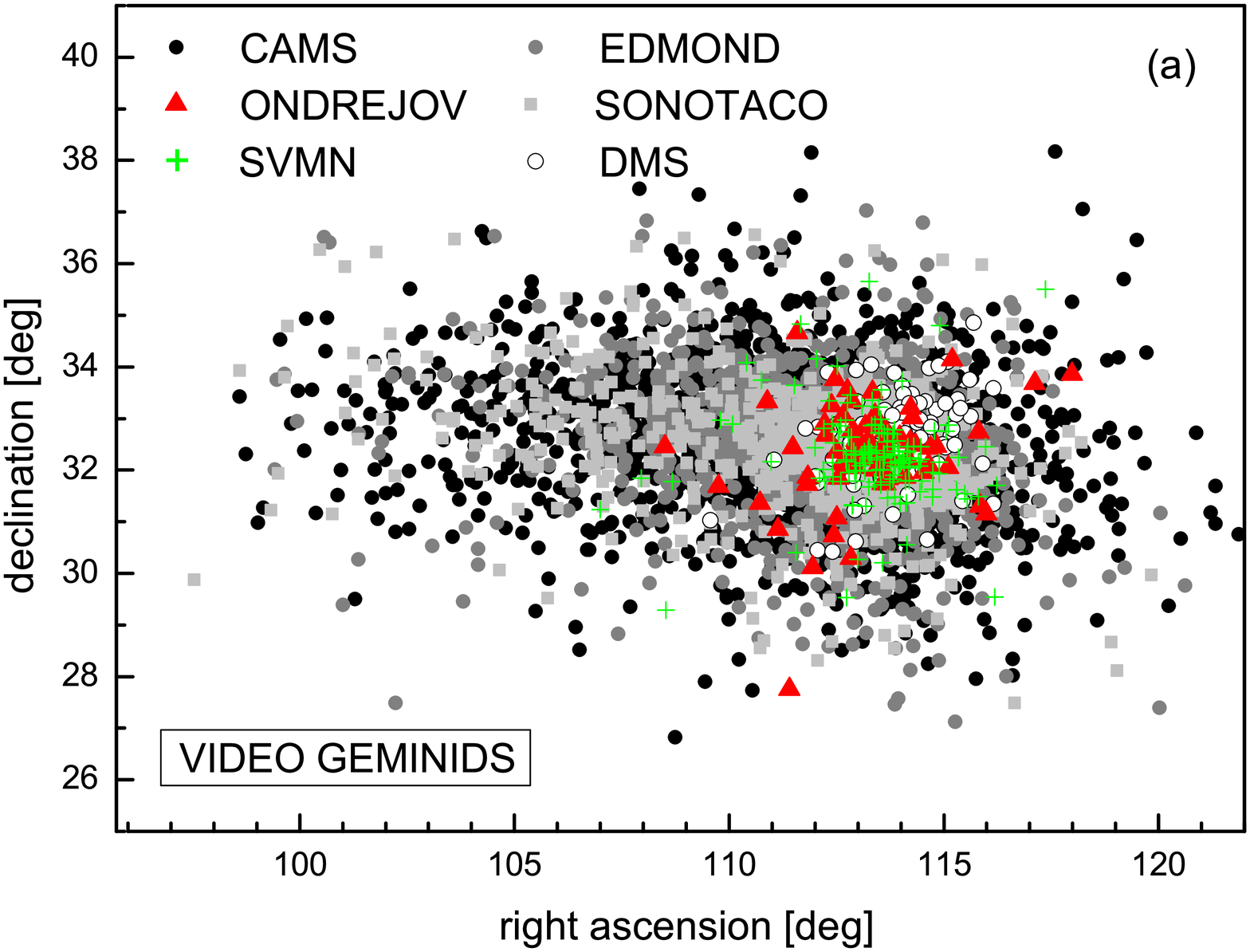}
            \includegraphics[width=6.8cm]{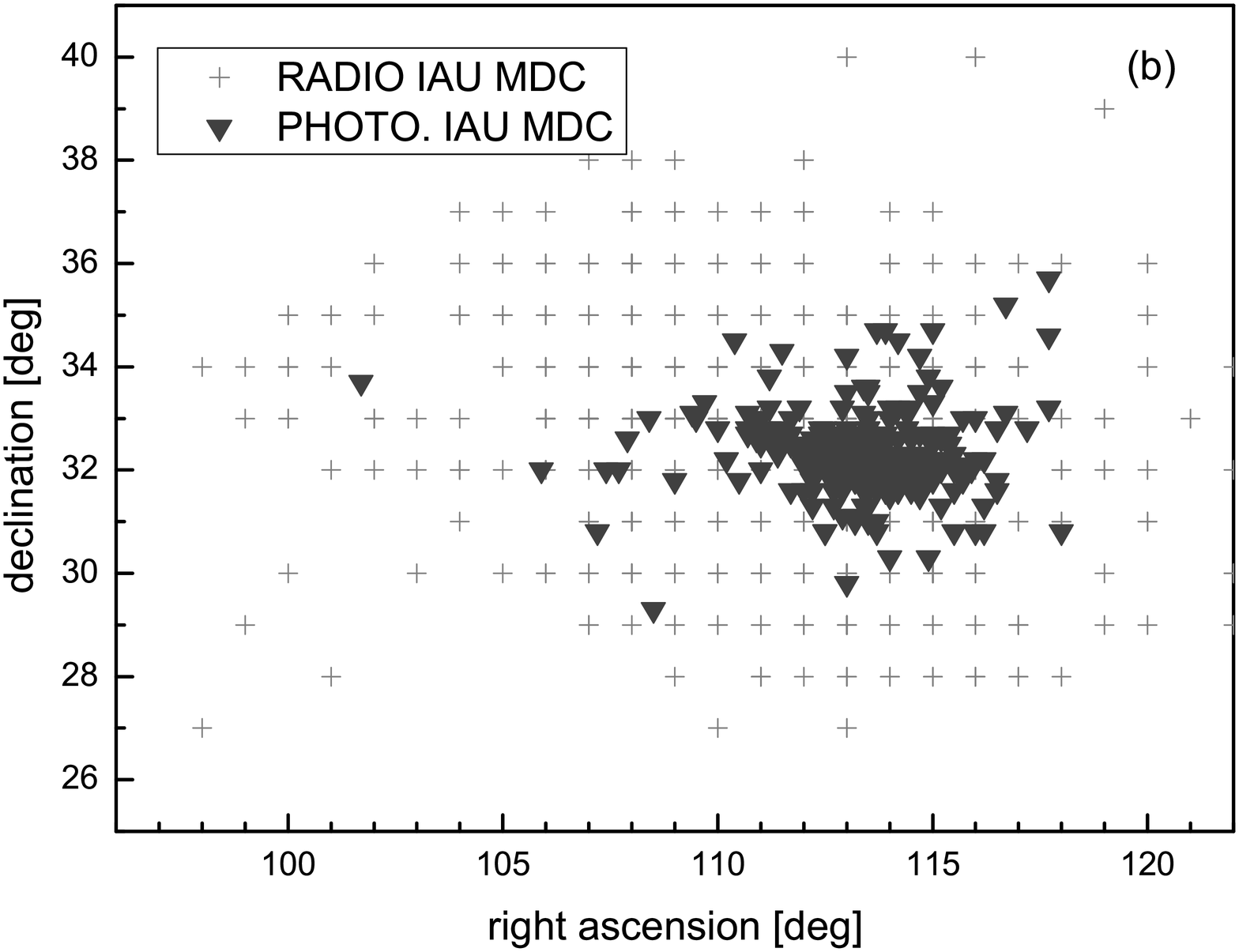}}
{\includegraphics[width=6.8cm]{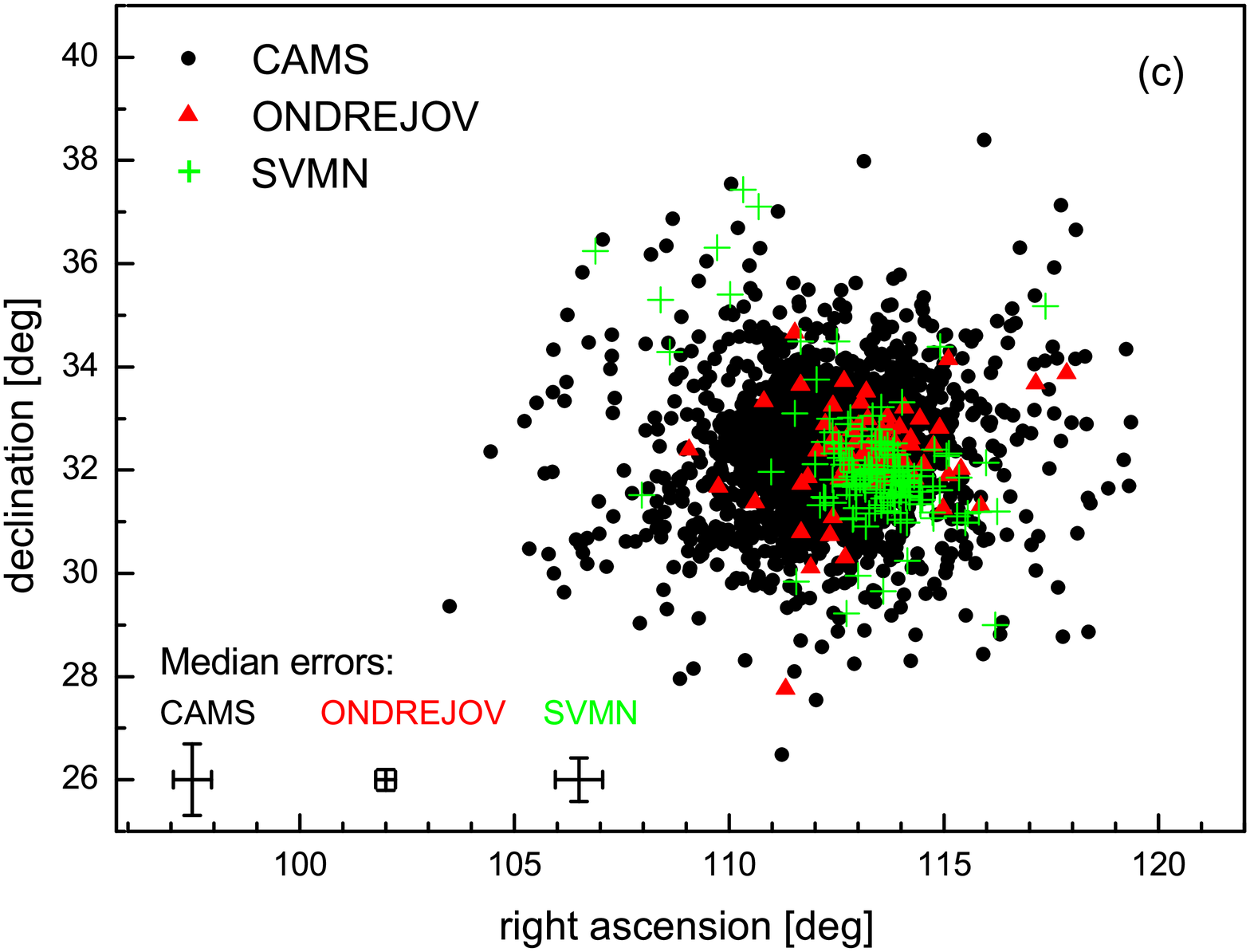}} \caption[f2]{(a) The
positions of the geocentric radiants of the Geminids obtained from
various video data and, for the sake of comparison, (b) from the
radio and photographic catalogues of the IAU MDC. (c) Radiant
points corrected for the radiant motion for Slovak, Czech and CAMS
Geminids.} \label{FIG6}
\end{figure}

The compactness of the Geminid stream is reflected in the
relatively small radiant area of the shower
\citep{kresakporubcan1970, porubcanetal2004}. The distributions of
the geocentric radiants of the Geminids from various data sets are
shown in figure 2a. The figure shows a good agreement in the
radiants among all the data, whereby large data sets show larger
dispersions, reaching the dispersion of the radio observed
Geminids (fig. 2b). Radiant points of the SVMN and the
Ond\v{r}ejov Geminids show the highest concentrations. However,
one would expect a wider spread in the SVMN data, taking into
account their given uncertainties (seen in the figure 1). This
suggests that the formal uncertainties in the SVMN data could be
overestimated.

Figure 2c shows the radiant positions corrected for the radiant
motion for Slovak, Czech and CAMS Geminids. The daily motion in
right ascension and declination was found by the least-squares
solution. The corresponding determined equatorial coordinates of
the mean geocentric radiants are listed in Table 2. The
$\mathrm{L_S}$ is the solar longitude of the time of observation
for equinox 2000.0. The solar longitude used for the radiant
corrections was the mean value determined from each catalogue
separately. The vast majority of SVMN Geminids are within a very
small interval of solar longitude; therefore, the daily motion of
the radiant declination was not possible to determine confidently
for this data.

The radiant area of Geminids (figure 2c) is approximately 15 x 10
deg obtained from the large CAMS catalogue and about 10 x 5 deg
obtained from the Ond\v{r}ejov data. The found spread includes the
dispersion caused by measurement errors. The calculated median
uncertainties of Czech/Ond\v{r}\v{e}jov (Slovak/SVMN) Geminids are
0.23 deg (0.55 deg) in right ascension and 0.16 deg (0.42 deg) in
declination. The corresponding values for CAMS data are 0.44 deg
and 0.69 deg. \cite{kresakporubcan1970} found the median real
dispersion of the photographic Geminids to be 0.49 deg, which
suggests how strong is the influence of measurement errors in the
video data.

\begin{table}[t] \large
\begin{center}
\caption{\it The radiant ephemeris for the video Geminids
determined from Slovak, Czech and CAMS databases} \label{t1}
\begin{tabular}{lll}
\hline
Database & right ascension (2000.0) &  declination (2000.0) \\
\hline
{\normalsize SVMN}        & $113.2 + 0.91(L_S - 261.8)$    & $ 32.3 - 0.02(L_S - 261.8)$ \\
{\normalsize OND\v{R}EJOV}    & $113.3 + 1.11(L_S - 262.1)$    & $ 32.3 - 0.11(L_S - 262.1)$ \\
{\normalsize CAMS}        & $112.5 + 1.02(L_S - 261.1)$    & $ 32.4 - 0.14(L_S - 261.1)$ \\
\hline
\end{tabular}
\end{center}
\end{table}

\subsection{Radiant dispersion as a function of semi-major axis}

To analyse the radiant area in connection with the distribution of
meteoroids with different semi-major axes, we divided the data
into three samples. The number of meteors in each sample was kept
near 1/3 of the total number. The results for four catalogues are
shown in Figure 3 a, b, c, d. The plots demonstrate that the
diffused marginal regions of the Geminid radiant are affected
mostly by meteoroids with extreme values (small or large) of the
semi-major axes. The radiant points of the meteoroids with
semi-major axes from the middle interval (which mostly define the
stream) are more concentrated and create a central area of about 5
x 5 deg obtained from the larger catalogues.

In all four data sets, meteoroids of shorter semi-major axes
occupy the eastern side of the radiant area, whereas those of
longer semi-major axes concentrate at the western part. A similar
displacement of radiant points, depending on the semi-major axes,
due to a vectorial composition of velocities, was found by
\cite{kresakporubcan1970}. Furthermore, the dispersion in the
right ascension seems to be slightly greater for lower values of
a, that in the declination, for higher values of a (seen
especially in the Ond\v{r}ejov data).

\begin{figure}
\centerline{\includegraphics[width=6.8cm]{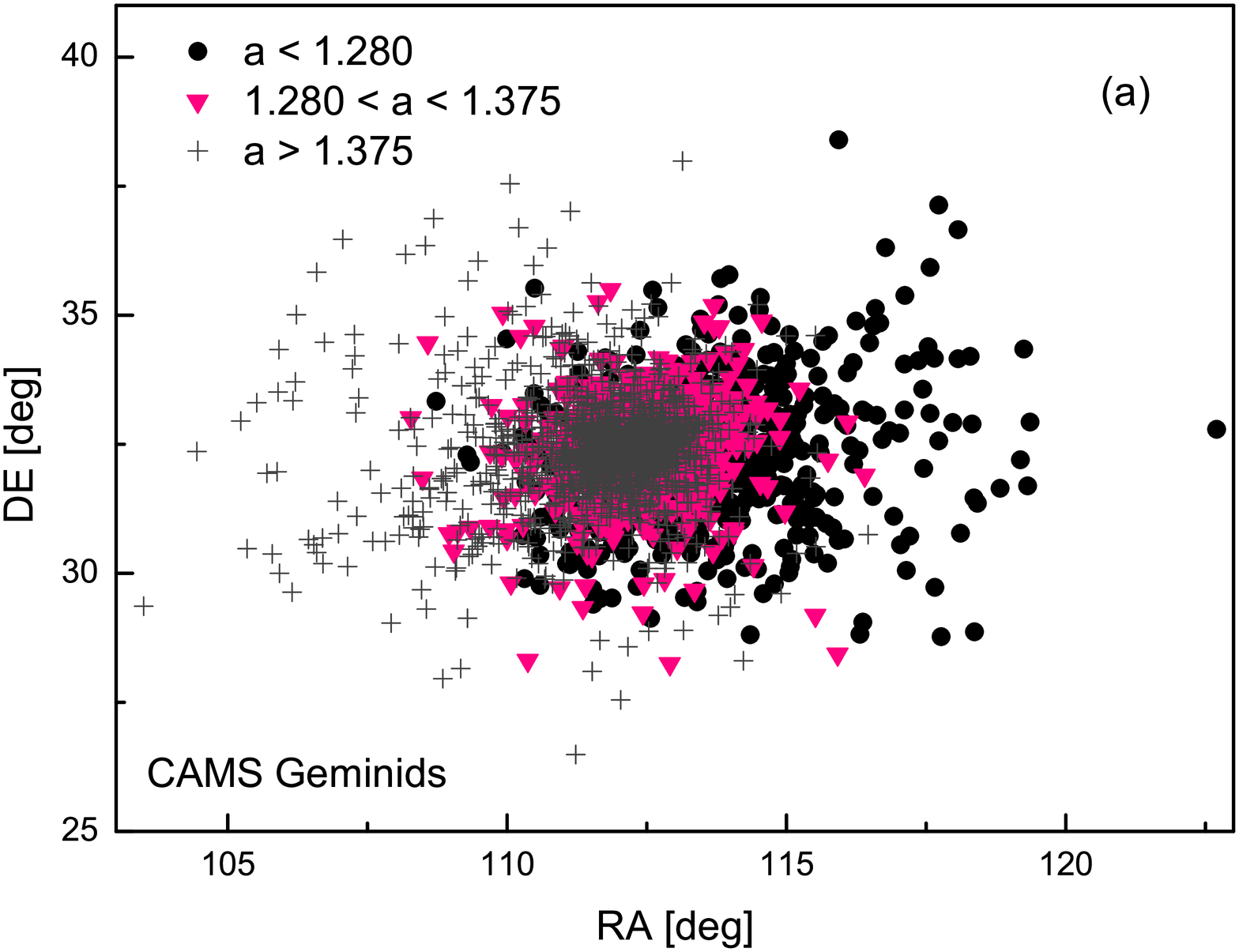}
            \includegraphics[width=6.8cm]{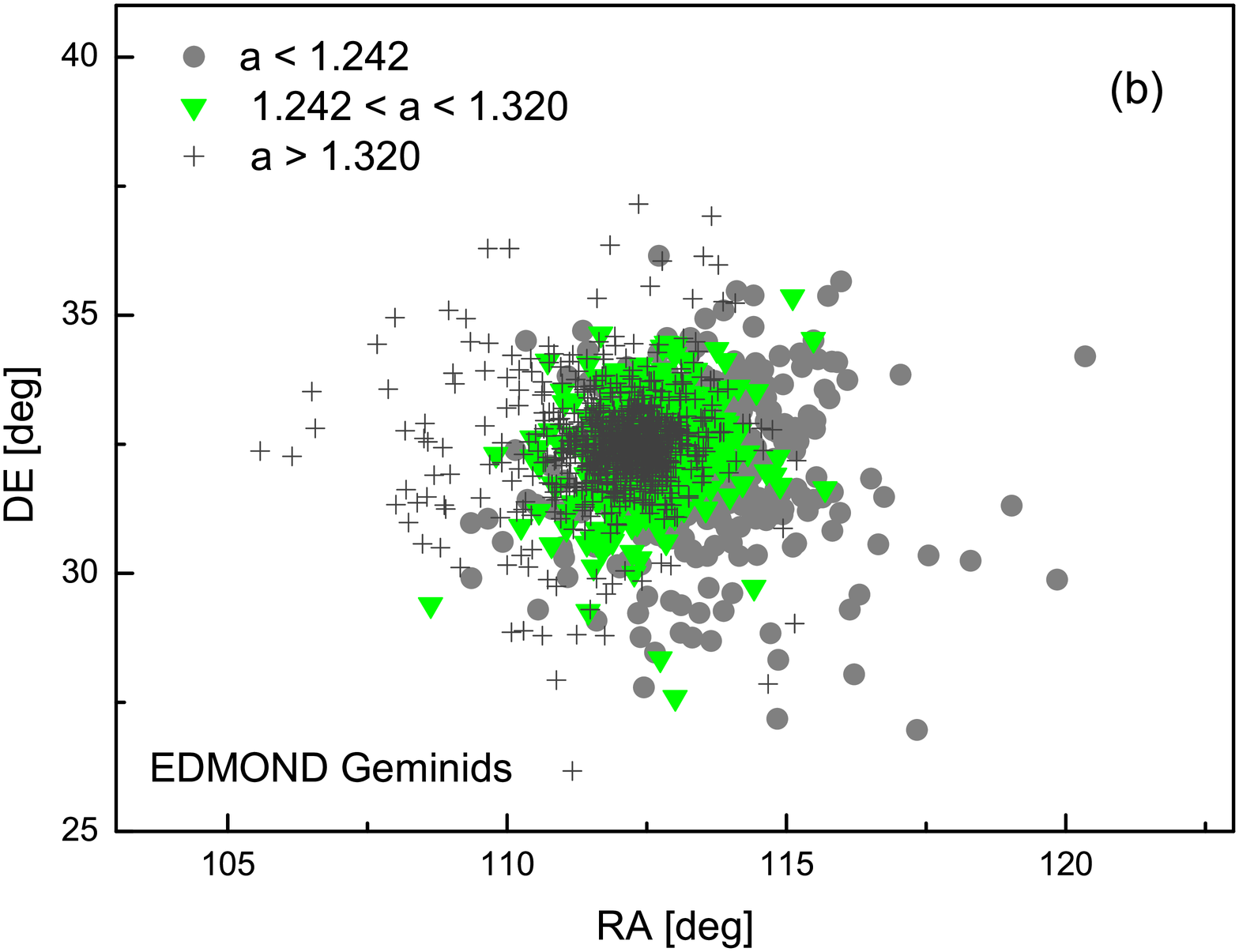}}
\centerline{\includegraphics[width=6.8cm]{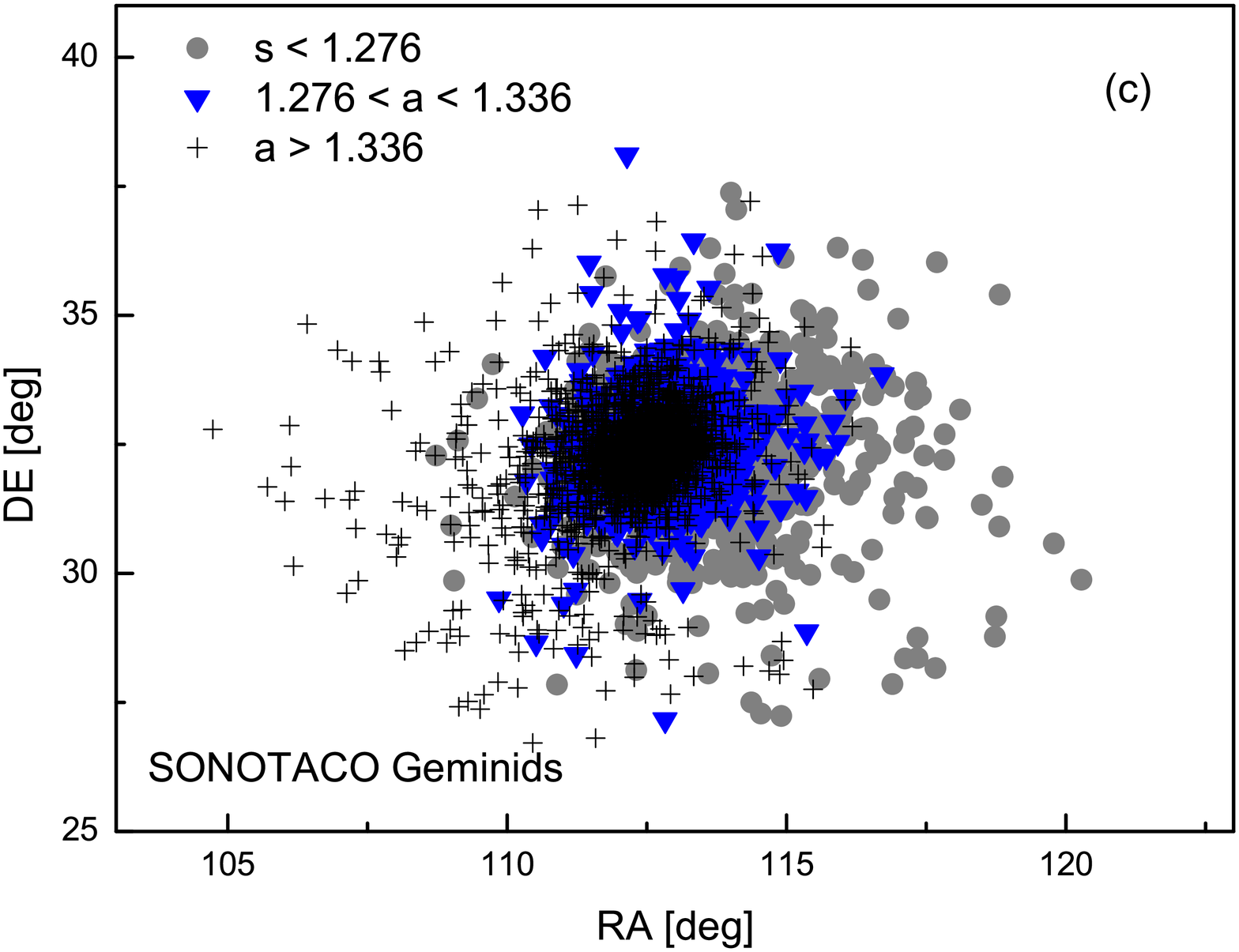}
            \includegraphics[width=6.8cm]{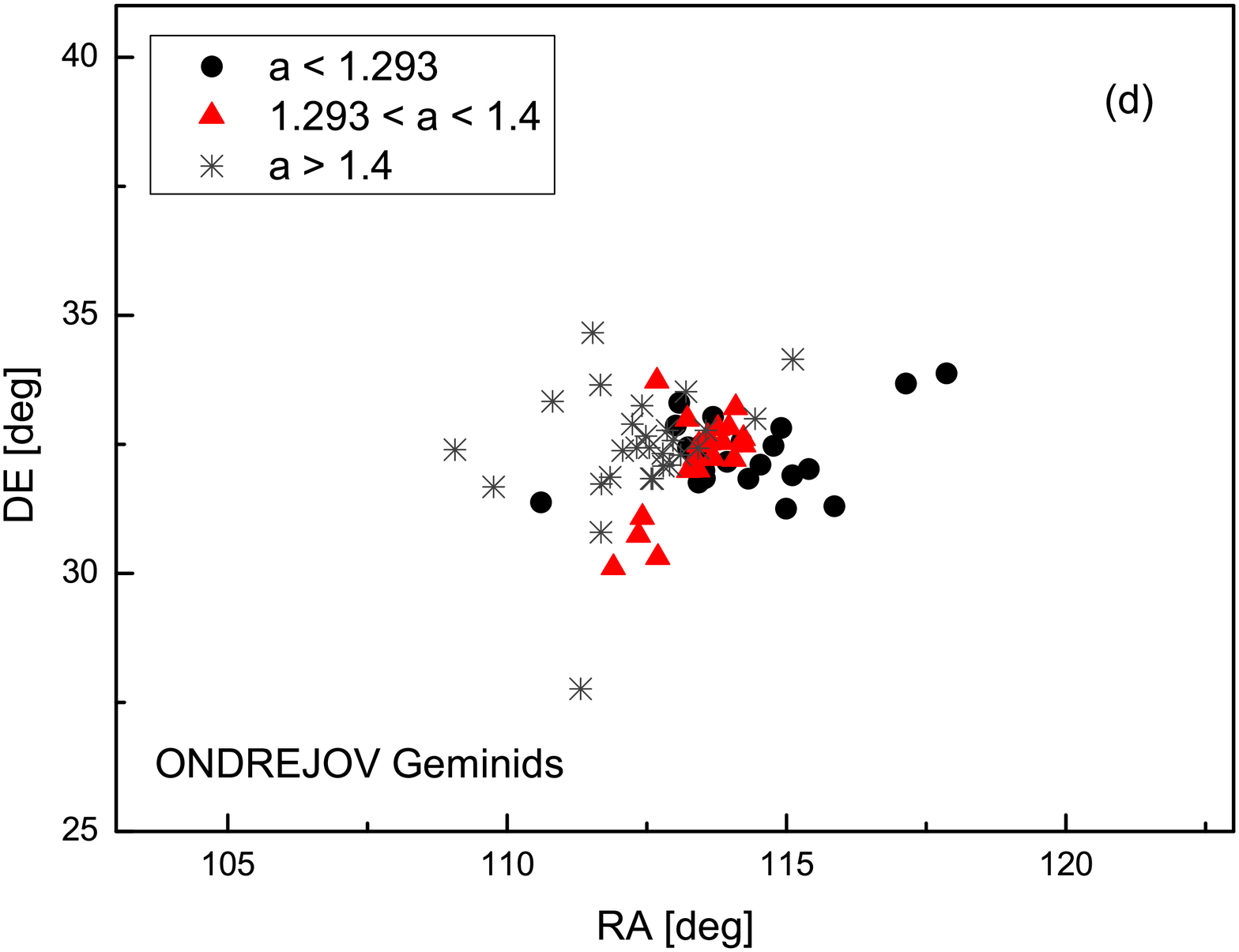}}
\caption[f2]{Radiant dispersion as a function of semi-major axis
for video Geminids from CAMS (a), EDMOND (b), SonotaCo (c), and
Ond\v{r}ejov (d) data. } \label{FIG6}
\end{figure}

\subsection{Velocity distribution - a shift in the video data}

\begin{figure}
\centerline{\includegraphics[width=6.8cm]{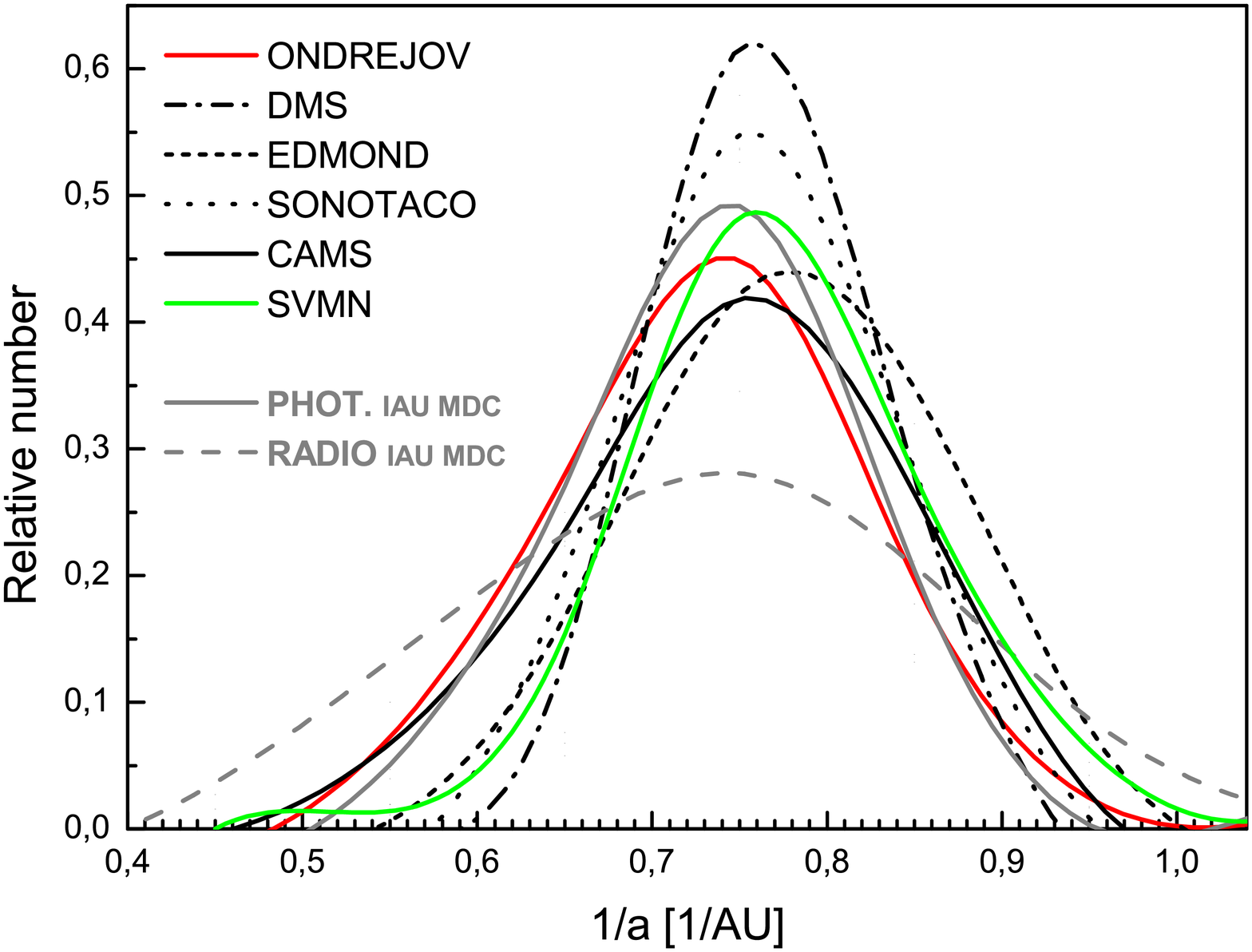}
            \includegraphics[width=6.8cm]{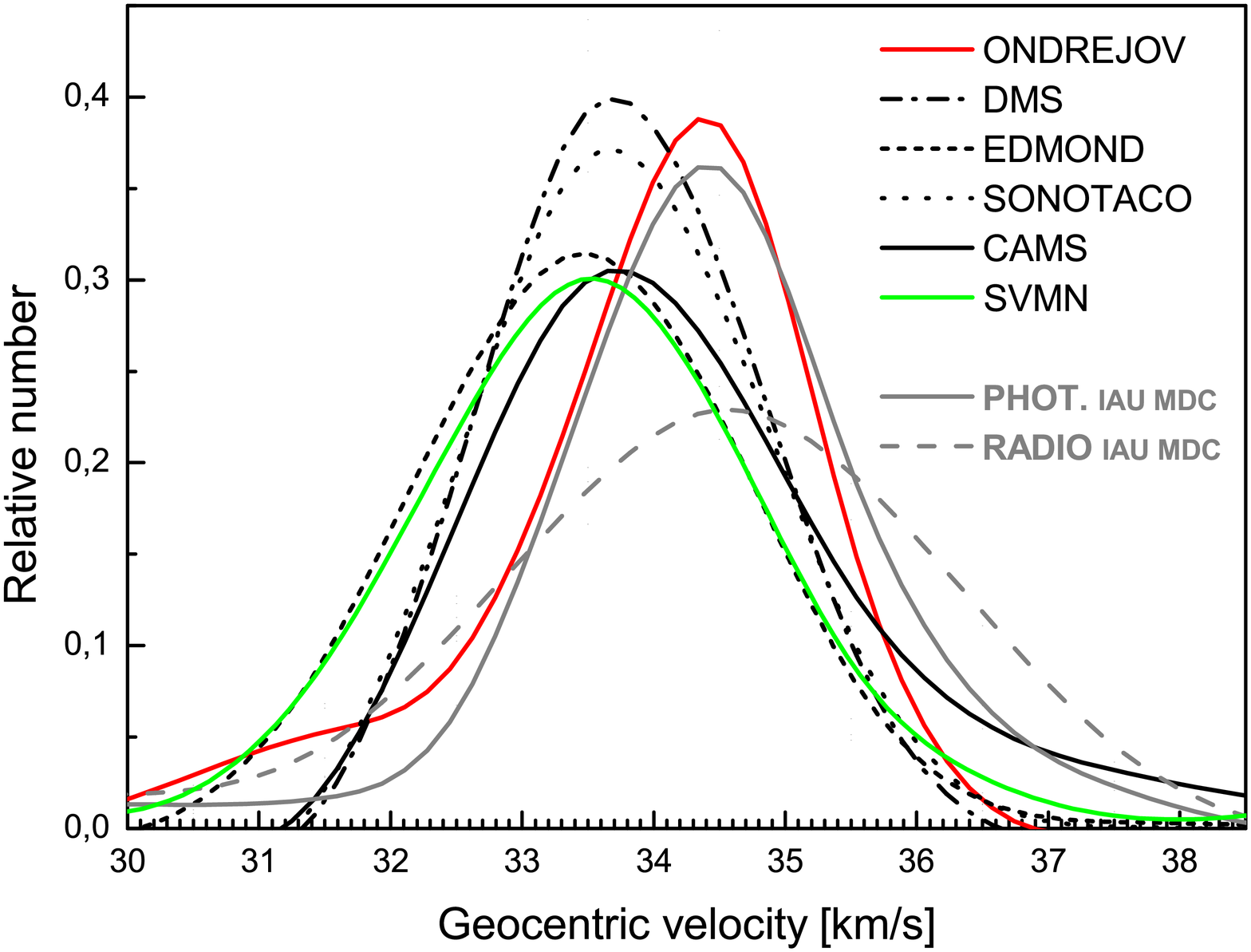}}
\caption[f2]{Normalised distributions of the reciprocal semi-major
axes and geocentric velocities of the video Geminids from the
different catalogues used, and their comparison with the
photographic and radio Geminids from the IAU MDC. The observed
shift in semi-major axes medians (and in the median values of the
geocentric velocity) from the video data is visible in almost all
the catalogues except for the Ond\v{r}ejov data.} \label{FIG6}
\end{figure}

Normalized distributions of the reciprocal semi-major axes and
geocentric velocities of the video Geminids from all of the
catalogues used, and their comparison with the photographic and
radio Geminids from the IAU MDC are shown in Figure 4 a and b. The
distribution of both the semi-major axes of meteor orbits and the
geocentric velocities in all the video data sets except the
Ond\v{r}ejov data seem to be systematically biased in comparison
with the photographic and radar meteors. The observed
distributions in $\mathrm{1/a}$ are shifted towards higher values
of $\mathrm{1/a}$. The determined velocities seem to be
underestimated, probably as a consequence of the methods used for
the measurement of the meteor positions and velocities (due to
measuring of the center of the meteor image, and absent or
insufficient correlations for atmospheric deceleration). The
largest shift is observed in the EDMOND and SonotaCo data, both
determining the speed value as an arithmetic mean. In the SVMN
database, it is possible to determine their entry velocity with a
higher precision, though only for the meteor data which allow us
to determine the deceleration using the exponential fit. The
velocities in the Czech data are computed from the manual
measurements of the individual meteor points. For each frame the
meteor position is measured not as the position of the head edge
of the meteor but the effect of blooming is taken into account.
Therefore, the position is measured "inside" the meteor image. The
distance from the edge is the same as the half-width of the meteor
image. The measurement of the image edge can lead to overvaluing
of the velocity, whereas the measurement
of the image center causes undervaluing.\\

\begin{figure}
\centerline{\includegraphics[width=6.8cm]{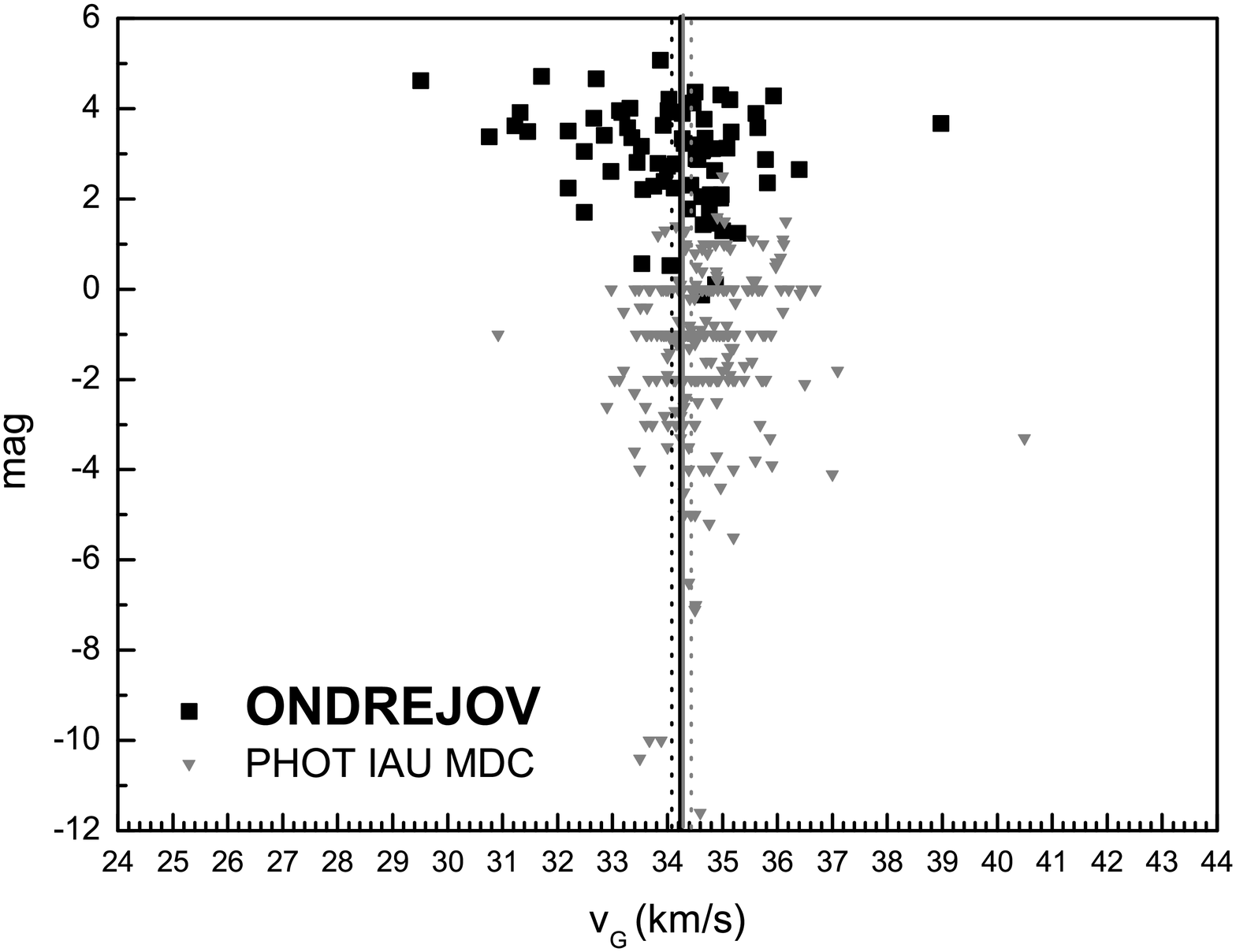}
            \includegraphics[width=6.8cm]{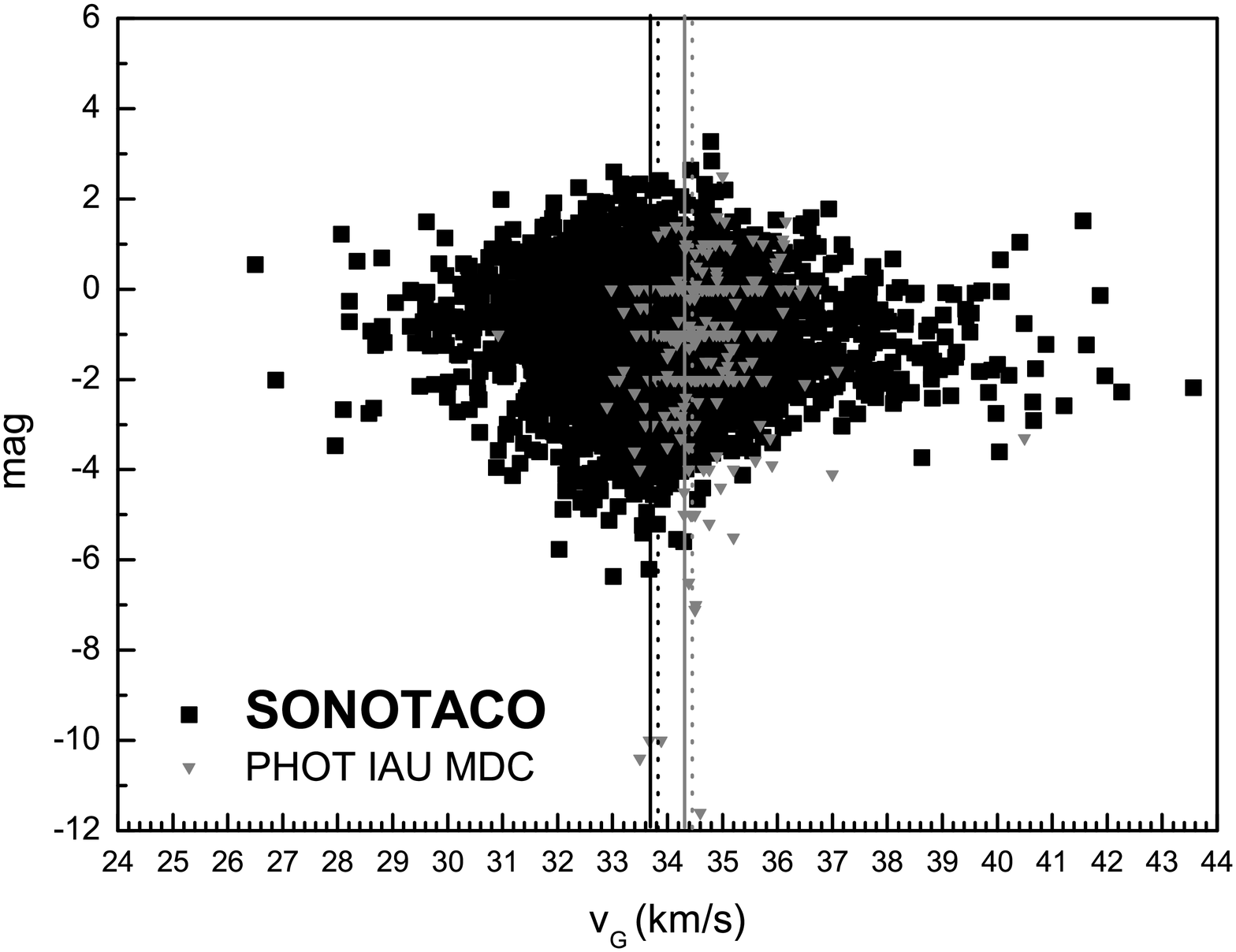}}
\centerline{\includegraphics[width=6.8cm]{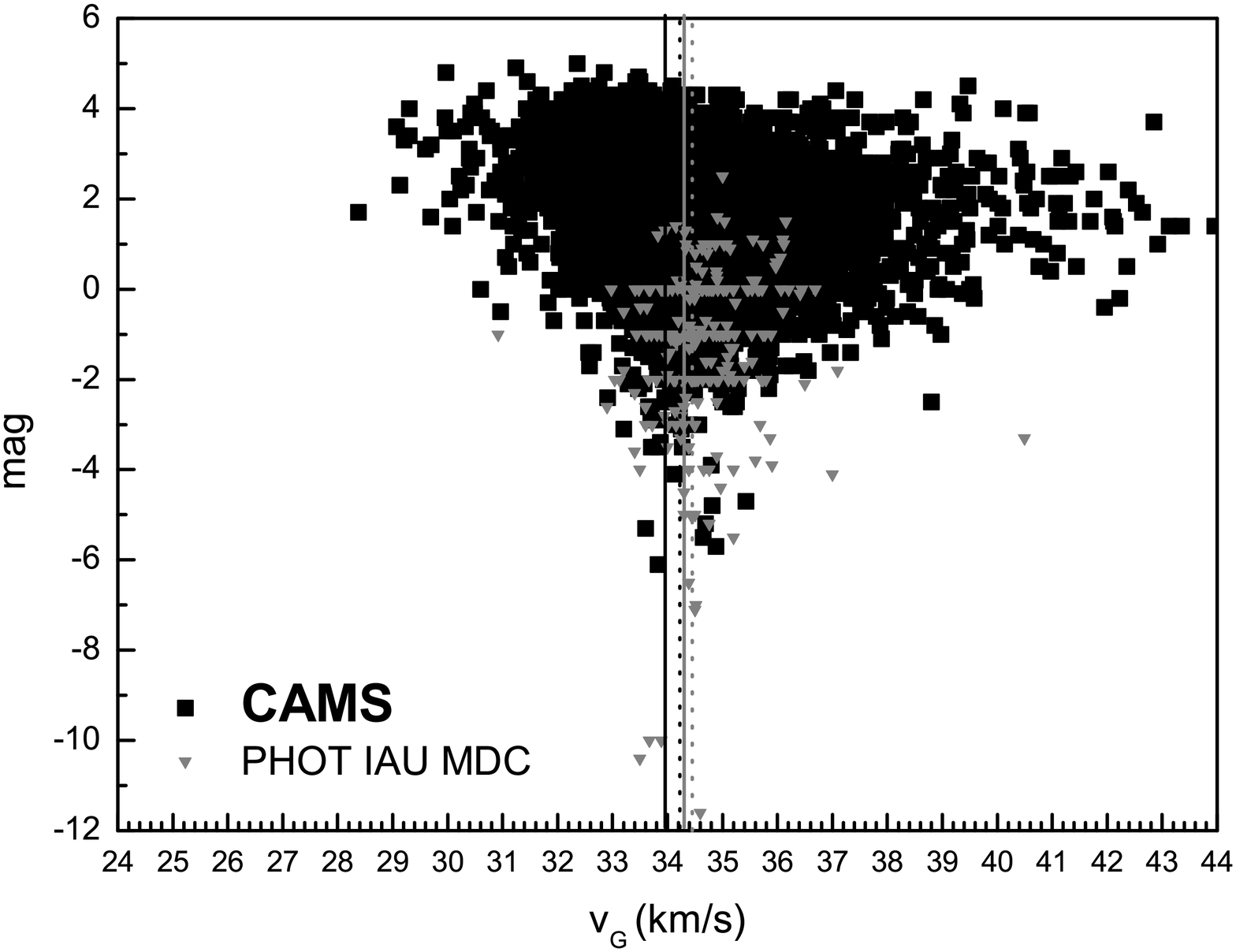}
            \includegraphics[width=6.8cm]{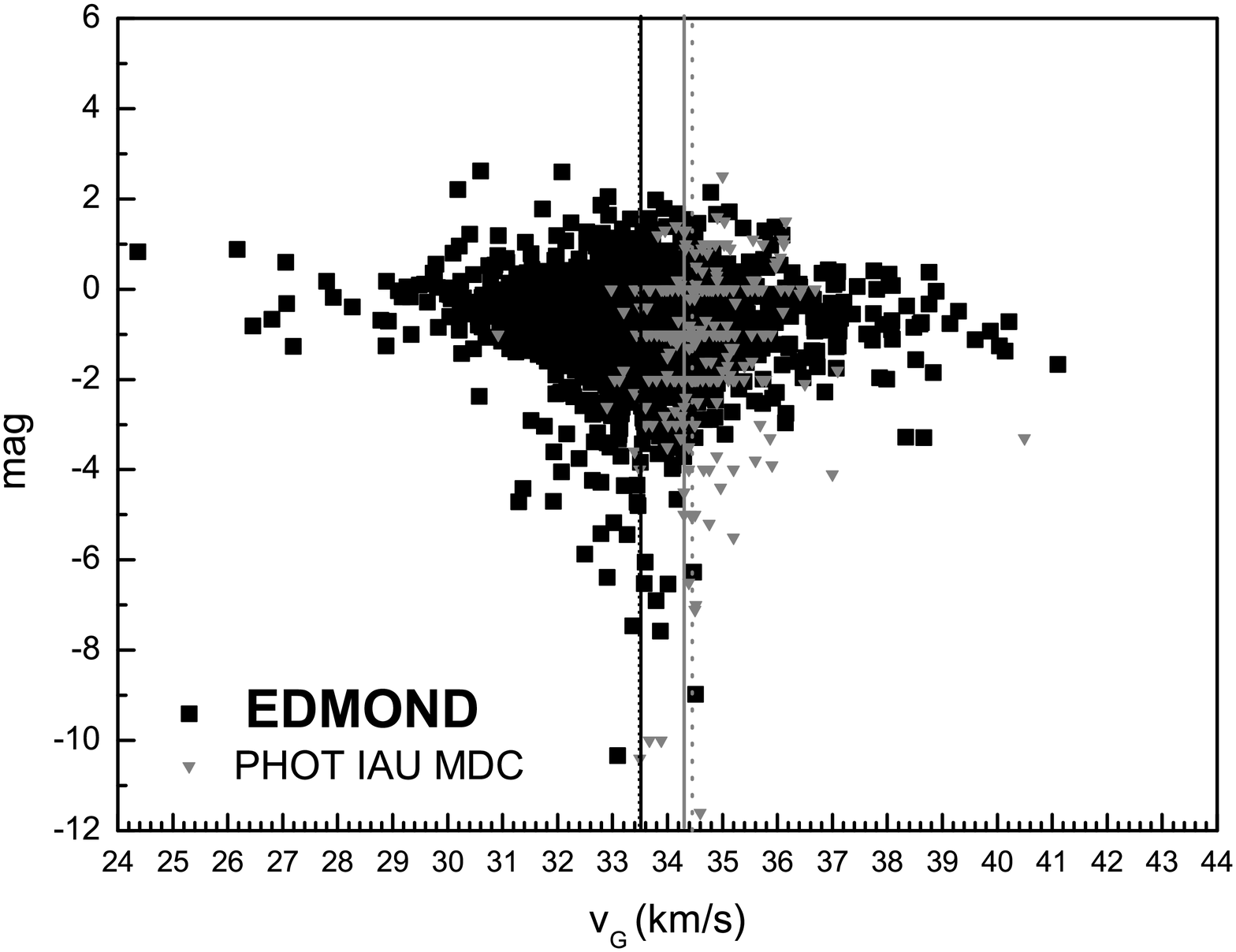}}
\caption[f2]{Magnitude as a function of geocentric velocity of the
Geminids from different video data sets (black squares), compared
with the photographic Geminids from the IAU MDC (grey triangles).
The vertical lines represent the mean (dotted line) and median
(solid line) values of the velocity determined from the video
(black) and photographic (grey) data.} \label{FIG6}
\end{figure}

To demonstrate the shift in the video data more clearly, we
plotted magnitudes of Geminid meteors as a function of their
geocentric velocity (figure 5). Each video data set (black
squares) is separately compared with the photographic Geminids
from the IAU MDC (grey triangles). The shift is demonstrated by
the gap between the vertical lines, showing the mean and median
values of the geocentric velocities from the video and
photographic catalogues. The smallest gaps are seen in the data
sets from Ond\v{r}ejov and CAMS, both registering fainter meteors
- of about 2 magnitudes more than SonotaCo or EDMOND observations,
the limiting magnitude of which is comparable with the
photographic data set. This fact shows that the shift is not
caused by the different masses of the data sets, but owing to the
errors. This is also supported by the comparison with the radar
observations registering even fainter meteors, distribution of
which does not show any shift in the semi-major axes in comparison
with the photographic data (see also fig. 4 and/or Table 3).

\section{Geminid stream. Orbits}

\subsection{Mean orbital parameters}

To compare all the various data, we selected, from all the data
sets, the Geminids that fulfilled the Southworth-Hawkins
D-criterion for orbital dissimilarity \citep{southworth1963} with
the condition $\mathrm{D_{SH}\,<\,0.2}$ (Table 1). We searched for
the core of the shower by comparing the density values of the
groups of meteors created by the Welch procedure \citep{welch2001}
around each orbit. The higher the density value, the more
important the group in the shower is. Consequently, we determined
the mean orbital elements and geocentric parameters of the densest
group by weighted arithmetic mean, using the Welch method. The
weight of the meteor was determined by $\mathrm{(1-D_i^2/D_c^2)}$
\citep{welch2001}. The resulted weighted mean parameters and their
standard deviations for each examined catalogue are listed in the
the Table 3. For a comparison, we listed also mean (not weighted)
parameters obtained from the radio and photographic catalogues of
the IAU MDC.

\begin{table}[t] \scriptsize
\begin{center}
\caption{\it The Geminids mean orbits determined from various
video data and compared with those of the photographic and radio
Geminids. For video data, the mean values of the orbital elements
and geocentrical parameters were determined by weighted arithmetic
mean, using the Welch method.} \label{t1} \hspace {0.1cm}
\begin{tabular}{llllllll}
\hline
Catalogue & $\mathrm{v_{G}}$ & a [AU] & e & q [deg] & i [deg]  & $\omega$ [deg]& $\Omega$ [deg] \\
\hline
OND\v{R}EJOV &  34.06$\pm$1.25 & 1.35& 0.15$\pm$.01 & 0.89$\pm$0.02  & 22.98$\pm$2.31 & 323.7$\pm$1.7 & 262.1$\pm$0.4 \\
SVMN &  33.55$\pm$1.37 & 1.29& 0.15$\pm$.01 & 0.89$\pm$0.02  & 22.42$\pm$1.92 & 324.1$\pm$1.7 & 261.9$\pm$1.3 \\
CAMS &  34.23$\pm$1.45 & 1.34 & 0.14$\pm$0.01 & 0.89$\pm$0.02  & 23.34$\pm$2.10 & 324.4$\pm$1.5 & 261.2$\pm$1.7 \\
SONOTACO &  33.81$\pm$0.01 & 1.30 & 0.15$\pm$0.01 & 0.89$\pm$0.01  & 22.85$\pm$1.71 & 324.2$\pm$1.3 & 261.5$\pm$1.6 \\
EDMOND &    33.49$\pm$1.21 & 1.28 & 0.15$\pm$0.01 & 0.88$\pm$0.02  & 22.49$\pm$2.01 & 324.3$\pm$1.5 & 261.2$\pm$1.6 \\
DMS &   33.80$\pm$0.88 & 1.31 & 0.14$\pm$0.01 & 0.89$\pm$0.01  & 22.35$\pm$2.09 & 324.1$\pm$1.0 & 262.3$\pm$0.1 \\
\hline
PHOTO$_{IAU}$       & 34.32 & 1.36 & 0.14 & 0.89 &  23.70 &  324.4 & 261.8 \\
RADIO$_{IAU}$      & 34.33 & 1.40 & 0.14 & 0.89 &  23.56 &  324.7 & 259.7 \\
\hline
\end{tabular}
\end{center}
\end{table}

\subsection{Orbital dispersion}

\begin{figure}
\centerline{\includegraphics[width=7cm]{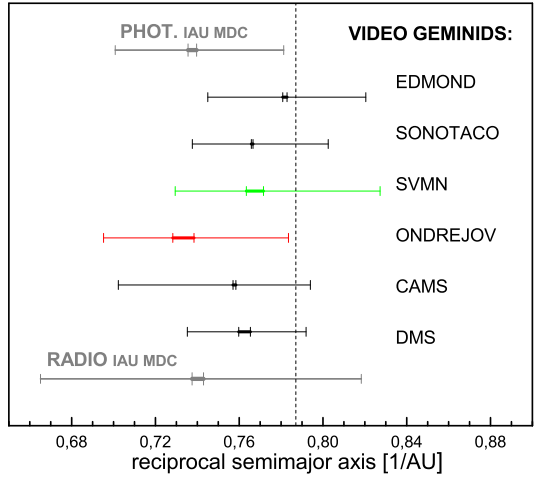}}
\caption[f2]{Observed orbital dispersion for Geminids described by
absolute median deviation in terms of $\mathrm{1/a}$: Thin line -
interval between two limiting values of $\mathrm{(1/a)_{1/2}}$,
which includes 50 percent of all orbits. Bold line - interval
between two limiting values of the uncertainty $\mathrm{(1/a)_L}$
of the resulting values of median $\mathrm{(1/a)_M}$. Dashed
vertical lines - parent body. Good agreement in the medians of
$\mathrm{1/a}$ is seen in the IAU MDC photographic, radio, and in
the Ond\v{r}ejov video data. In all other video databases, it is
shifted towards higher values.} \label{FIG6}
\end{figure}

The orbits of the Geminids with aphelia far inside the orbit of
Jupiter indicate that the stream structure is dominated by their
initial spread and the non-gravitational effects
\citep{williamsryabova2011}. However, the gravitational forces of
Jupiter and inner planets also influence the stream structure
\citep{ryabova2014}, while the other outer planets influence is
negligible. The differences in the velocities are less
representative, and the dispersion in the semi-major axes smaller
in comparison with long-period streams.

In analyzing the error function, we proceeded according to the
analysis of the accuracy of the semi-major axes of meteoroid
orbits made by \cite{kresakova1974}. To describe the dispersion of
the semi-major axis within the meteor stream, we used the median
absolute deviation $\mathrm{\Delta_M}$ in terms of $\mathrm{1/a}$
($\mathrm{\Delta_M(1/a)\,=\,\mid(1/a)_{1/2} - (1/a)_M \mid}$,
where $\mathrm{(1/a)_{1/2}}$ are limiting values of the interval,
which includes 50 percent of all orbits in the stream). The
probable range of uncertainty was determined by $\mathrm{\pm
n^{-1/2}\Delta_M(1/a)}$, where n is the number of the meteor
orbits used for the median determination $\mathrm{(1/a)_M}$. For
the sake of comparison, we also derived the deviations of the
median $\mathrm{(1/a)_M}$ from the parent body
$\mathrm{\Delta(1/a)_{C}\,=\,\mid(1/a)_M - (1/a)_{C}\mid}$, where
the $\mathrm{(1/a)_{C}}$ is the reciprocal semimajor axis of
asteroid (3200) Phaethon. The numerical results describing the
observed orbital dispersion within the video Geminids are listed
in Table 4. Their comparison with the Geminids' dispersion from
the photographic and radar data is shown in figure 6. The median
absolute deviations $\Delta_M$ in terms of $\mathrm{1/a}$ range
from $0.028$ to $\mathrm{0.049\,AU^{-1}}$ for the different video
catalogues. The figure shows the dispersion of video Geminids to
be comparable with the dispersion of photographic Geminids
($\mathrm{0.040\,AU^{-1}}$), and to be significantly smaller than
the dispersion of radio Geminids ($\mathrm{0.077\,AU^{-1}}$). This
is partly a consequence of different dispersions in the orbital
elements for particles belonging to different mass ranges. The
limiting magnitude of the SVMN, SonotaCo or EDMOND observations is
comparable with that of the photographic data set, while that of
the Ond\v{r}ejov, CAMS and DMS data, is about 2 magnitudes more;
and the radio observations register even fainter meteors. It also
has to be considered that the video data were observed mostly
during the last few years, whereas the photographic data cover
more than 30 years of observations; thus, we are dealing with
particles from a different cross section of the stream traversed
by the Earth.

The deviation of the median reciprocal semi-major axis from the
parent body, asteroid (3200) Phaethon, obtained from the
photographic and radar orbits of the IAU MDC, and from the Czech
Video Orbits Catalogue, is significantly larger than in the other
video data sets. However, this is only a consequence of their
above-mentioned shift. The actual reason for the deviation from
their parent asteroid might be found when investigating the
dynamical evolution of the Geminid meteoroids and the 3200
Phaethon.

\begin{table}[t] \small
\begin{center}
\caption{\it Numerical data obtained for the orbital dispersion of
the video Geminids. n - number of meteors; $\mathrm{(1/a)_M}$ -
the median $\mathrm{1/a}$; $\mathrm{1/a}$ - the mean value of
$\mathrm{1/a}$; $\mathrm{\Delta_M(1/a)}$ - the median absolute
deviation; $\mathrm{\Delta_L(1/a)}$ - the range of uncertainty;
$\mathrm{\Delta(1/a)_C}$ - deviation of the median $\mathrm{1/a}$
from the parent body; $\mathrm{1/a}$ of the parent comets (eq.
2000.0).} \label{t1} \hspace {0.1cm}
\begin{tabular}{lrrrrrrr}
\hline
Database    & No of & median      & median      & mean    &  &  & \\
        & orbits         & $\mathrm{v_{G}}$ & $\mathrm{(1/a)_M}$ & $\mathrm{1/a}$  & $\mathrm{\Delta_M(1/a)}$ & $\mathrm{\Delta_L(1/a)}$  & $\mathrm{\Delta(1/a)_C}$\\
\hline
{\footnotesize SVMN}        & 143   & 33.68 & 0.766 & 0.773     & 0.049  & 0.004    & - 0.021\\
{\footnotesize OND\v{R}EJOV}    & 74    & 34.29 & 0.734 & 0.775     & 0.044  & 0.005    & - 0.050\\
{\footnotesize CAMS}        & 4829  & 33.97 & 0.758 & 0.739     & 0.046  & 0.001    & - 0.029\\
{\footnotesize SONOTACO}    & 8264  & 33.82 & 0.766 & 0.764     & 0.036  & 0.001    & - 0.021\\
{\footnotesize EDMOND}      & 2401  & 33.50 & 0.781 & 0.780     & 0.038  & 0.001    & - 0.006\\
{\footnotesize DMS}     & 104   & 33.80 & 0.762 & 0.761     & 0.028  & 0.003    & - 0.025\\
\hline
\end{tabular}
\end{center}
\end{table}
\hspace {0.5cm}

\section{The Geminid-stream dynamics and the uncertainties}

The main purpose of this paper is to emphasize the problem of
measurement/determination errors, which have to be considered when
interpreting results. Since similar problem also applies to
analyzes based on numerical integration, we include a short
section about the dynamics of the stream. We would like to briefly
point out the problem of errors in the integration procedures
which are used for stream modeling. The uncertainties influence
the outputs of the simulations and give us important information
about the reliability of the results obtained.

\subsection{Models versus observations}

The initial dispersion of meteoroids in a stream is influenced by
a number of processes (planetary perturbations, collisions, solar
radiation, solar wind and other non-gravitational forces), which
appear during different stages of the stream evolution.
\cite{williamsryabova2011} have discussed their influence on the
structure of meteoroid streams, and demonstrated that the dominant
process depends on the stream and thus, for the stream models, it
is important to consider both the initial processes of formation
and the subsequent gravitational perturbations. There have not
been any close encounters significantly affecting the Geminids'
orbits during at least the last ten thousand years
\citep{ryabova2007}, so the initial structure caused by the
ejection process should still be traceable in the stream. The
deviations which may have accumulated since the formation of the
stream can hardly exceed a few thousandths in $\mathrm{1/a}$
\citep{kresakova1974}. Nevertheless, the non-gravitational forces
have an influence on both the orbits of the meteoroids (e.g. P-R
drag) and the orbit of the asteroid (e.g. rocket effect/yet force)
\citep{galushinaetal2015, ryabova2016}.

Over time, various Geminid stream models have been developed
\citep{foxetal1982, foxetal1983, foxetal1984, williamswu1993,
babadzhanovobrubov1983, babadzhanovobrubov1986,
babadzhanovobrubov1987, ryabova2001, ryabova2007, ryabova2008,
ryabova2016, kanuchovasvoren2006}. A compilation of them can be
found in \cite{neslusan2015} and their detailed individual
significance explained in \cite{ryabova2014}. In spite of the
large number of Geminid stream studies, there are still
discrepancies between the models and observations, e.g. the width
of the stream, the location of the stream and the maximum of the
shower activity \citep{ryabova2016}. The Geminid shower width does
not increase, even if encounters with the inner planets are
included into consideration in models (Ryabova, 2016). The most
probable reason for the discrepancies is the transformation of the
parent body orbit by the jet force \citep{ryabova2014,
ryabova2016} according to the \cite{lebedinets1985} hypothesis of
a rapid release of the volatiles in the process of the stream
single initial formation.

In general, if the orbit of a parent body was, in the past,
significantly influenced by non-gravitational effects, the
currently observed stream could have been formed when the parent
body moved in a slightly different orbit than it moved at its last
return to the perihelion \citep{neslusanetal2015}. However, the
Phaethons' semi-major axis is stable, and the changes in the
asteroid's other orbital elements are smooth \citep{ryabova2008,
jakubikneslusan2015}.

\cite{jonesetal2016} argued that most attempts to model the
Geminid meteor stream had been based on Whipple's model for the
ejection of meteoroids from comets (assuming ejection speeds about
a factor of at least 3 too low), which predicts much smaller
dispersions in terms of the orbital elements than are found in the
observed behaviour of the Geminids. However,
\cite{jakubikneslusan2015} showed in their simulations that no
matter if they assumed the unique speed and random directional
distribution of the test particles or a more realistic ejection,
the appropriate orbital phase space was fullfiled with test
particles after a certain period (which is, using their model,
about 30 orbital revolutions of the parent). This means, in the
case of the asteroid 3200, it is less than a century. And most of
the simulations exceed this time.

The authors \citep{jakubikneslusan2015} also examined the
acceleration due to the P-R drag on the Geminid test particles
which should, in general, cause an enlargement of the dispersion
of orbits. This, however, may not be reflected in the
corresponding meteor shower observed in the Earth's atmosphere,
since some stream particles can be completely deflected from a
collisional course with the Earth. According to their model, the
physical properties of the prevailing part of the Geminids
correspond to the values from 0.005 to 0.018 of the $\beta$
parameter, which determines the strength of the P-R drag. The beta
parameter stands for the ratio of the radiation pressure to the
gravity, and characterizes the properties of the particle; thus,
only the Geminids with the corresponding properties can be
observed.

\subsection{The simulations reliability}

\begin{table}[t] \normalsize
\begin{center}
\caption{\it The actual orbit of Phaethon and its orbits after the
integration to the past for 10\,000, 20\,000 and 50\,000 years and
then back to the present. Differences in particular orbital
elements can reach the following values: $\Delta a<$ 0.01\,AU,
$\Delta e \sim$ 0.003, $\Delta i \sim 1.5^\circ$, $\Delta \omega
\sim 3.0^\circ$, $\Delta \Omega \sim 3.5^\circ$ and $\Delta q
\sim$ 0.004\,AU.} \label{t1} \hspace {0.1cm}
\begin{tabular}{rllllll}
\hline
                & a [AU]    & e         & i [deg]      & peri [deg]   &  node [deg] & q [deg] \\
\hline
present         & 1.2711    & 0.8898    & 22.24        & 322.14  &  265.27 & 0.1400 \\
after 10000 yr  & 1.2740    & 0.8897    & 22.19        & 322.05  &  265.29 & 0.1404 \\
        20000 yr   & 1.2780    & 0.8882    & 22.94        & 323.67  &  263.04 & 0.1428 \\
        50000 yr   & 1.2772    & 0.8870    & 23.46        & 325.01  &  261.80 & 0.1443 \\
\hline
\end{tabular}
\end{center}
\end{table}

We studied the evolution of the orbit of the asteroid (3200)
Phaethon by means of numerical integration using Everhart's
integrator RA15, from the package Mercury 6 \citep{chambers1999}.
The model of the Solar System used in the integration tests
included 8 planets, the Moon as a separate body, and the most
influential asteroids: Ceres, Pallas, Vesta, and Hygiea
\citep{galad2001}.

Several experimental integrations of the asteroid, performed from
the present to the past and then back to the year 2015, showed
that it is not possible to reproduce the initial asteroid's orbit.
A complete reproduction, also including the mean anomaly, is only
possible for a time span of about 2700 years; the mean anomaly
differs only $\mathrm{3\,x\,10^{-6}\,deg}$. After 3800 years, the
difference is $\mathrm{\sim0.03\,deg}$ and after 4700 years, it is
not possible to reproduce the mean anomaly. However, the chaos
indicators behavior has to be taken into account. E.g. MEGNO (Mean
Exponential Growth of Nearby Orbits) shows that the chaos along
the Geminid's orbit begins after already 400 years of integration
\citep{galushina2016}, and Lyapunov indicator shows that it begins
after 1500 years approximately \citep{avdyushev2016}.

In addition, the results of the integration depend not only on the
chosen integrator, but also, within a particular integrator, on a
specific perihelion passage, on the selected accuracy parameter,
and on predetermined intervals of the outputs, which forces the
integrator to modify its own integration steps to match the output
moments.

\section{Summary and Conclusions}

We examined the influence of the uncertainties of the velocity
measurements and orbit determination on the dispersion of radiants
and meteoroid orbits within the stream of Geminids. The dispersion
was studied, comparing several databases, based on various
meteor-detection software packages and various meteor orbital
element softwares. For the analysis, data from our own video
observations, carried out in the Slovak and Czech republics, as
well as data from several avaliable video databases were used: the
Slovak Video Meteor Network's database \citep{tothetal2015}, the
Czech Catalogue of Video Meteor Orbits \citep{kotenetal2003}, the
CAMS Meteoroid Orbit Database \citep{jenniskensetal2011}, the
Dutch Meteor Society Video Database \citep{delignie1996}, the
SonotaCo Shower Catalogue \cite{sonotaco2009}, and the EDMOND
Database \citep{kornosetal2014}. The comparison of the errors
distribution was based on the velocity and radiant uncertainties
declared by the authors of the databases. To avoid the distortion
by extreme deviations caused by gross errors, we chose the medians
rather than arithmetic means as the basic dispersion parameter.
For the mean orbit determination, we used the weighted values of
the orbital elements.

The Geminids' radiant dispersion obtained from the large video
catalogues reaches the dispersion of the radio observed Geminids,
whereby the diffused marginal regions are affected mostly by
meteoroids with extreme values (small or large) of the semi-major
axes. Meteoroids of shorter semi-major axes occupy the eastern
side of the radiant area and those of longer semi-major axes the
western part.

The observed orbital dispersions in the Geminid stream described
by the median absolute deviation range from 0.028 to 0.049
$\mathrm{AU^{-1}}$ for the different video catalogues. It does not
differ significantly between the different video databases. It
differs slightly between the data sets obtained by different
observational techniques, which may be partly a consequence of
different dispersions in the orbital elements for particles
belonging to different mass ranges.

The distribution of the semi-major axes of video meteors in all
the video data sets, except for the Ond\v{r}ejov data, seem to be
systematically biased in comparison with the photographic and
radar meteors. The reciprocal semi-major axes are shifted towards
higher values; the determined velocities are underestimated. This
is a consequence of the methods used for the measurement of the
meteor positions and the velocity. The measurement of the image
edge can lead to overvaluing of the velocity, whereas the
measurement of the image center causes undervaluing. The largest
shift is observed in the EDMOND and SonotaCo data. The velocities
in the Czech data, which did not show any shift, are computed from
the manual measurements of the individual meteor points. For each
frame, the position is measured "inside" the meteor image. This
approach comes at the expense of the total number of trajectories
and orbits; however, the database contains only reliable data.

To avoid entering poor quality data into the databases, an
improvement in the measurement of the meteor positions and meteor
velocities is essential. Moreover, stronger filters for the
selection of orbits should be used, which will come at the expense
of the quantity of the data. However, for some purposes, only high
accuracy data are applicable, failing which, each analysis using
the velocity data will be seriously affected by measuring errors.
\\

Aside from those measurement errors which have an effect on the
analyses, we have also indicated the difficulties associated with
the uncertainties of the numerical integration procedures that
influence the results of the simulations. We performed several
experimental integrations of the Geminids' parent asteroid, from
the present to the past and then back to the year 2015, and
demonstrated that a complete reproduction is only possible for a
time span of about 2700 years. This finding impacts analyses based
on numerical integrations and/or their interpretations.


\hspace {0.5cm}\hspace {0.5cm}

\textbf{Acknowledgements.} We gratefully thanks Galina Ryabova for
helpful discussions and comments that have improved the article.
The work was supported, by the Slovak Grant Agency for Science
VEGA, grant no. 1/0225/14, by the Slovak Research and Development
Agency under the contracts no. APVV-0517-12, and by the Grant
Agency of Czech Republic, grant no. 14-25251S.


\hspace {0.5cm}\hspace {0.5cm}

\bibliographystyle{elsarticle-harv}
\bibliography{<your bibdatabase>}

\textbf{References}\label{}

\end{document}